# Reduced interhemispheric functional connectivity of children with autism: evidence from functional near infrared spectroscopy studies

Huilin Zhu,[1,2] Yuebo Fan,[3] Huan Guo,[2] Dan Huang,[3] and Sailing He,[1,4,*]

[1]*Centre for Optical and Electromagnetic Research, ZJU-SCNU Joint Research Center of Photonics, South China Normal University (SCNU), Guangzhou, 510006, P. R. China*
[2]*School of Psychology, South China Normal University (SCNU)), Guangzhou 510631, P.R. China*
[3]*Guangzhou Rehabilitation and Research Center for Children with ASD, Guangzhou, 510540, P.R .China*
[4]*JORCEP, Department of Electromagnetic Engineering, School of Electrical Engineering, Royal Institute of Technology, Stockholm, 10044, Sweden*
*\*sailing@kth.se*

**Abstract:** Autism spectrum disorder is a neuro-developmental disorder characterized by abnormalities of neural synchronization. In this study, functional near infrared spectroscopy (fNIRS) is used to study the difference in functional connectivity in left and right inferior frontal cortices (IFC) and temporal cortices (TC) between autistic and typically developing children between 8-11 years of age. 10 autistic children and 10 typical ones were recruited in our study for 8-min resting state measurement.  Results show that the overall interhemispheric correlation of HbO was significantly lower in autistic children than in the controls. In particular, reduced connectivity was found to be most significant in TC area of autism. Autistic children lose the symmetry in the patterns of correlation maps. These results suggest the feasibility of using the fNIRS method to assess abnormal functional connectivity of the autistic brain and its potential application in autism diagnosis.

**1. Introduction**

Optical methods have been demonstrated as a non-invasive way to probe human brain activities [1, 2]. Functional near infrared spectroscopy (fNIRS), which measures the absorption of near infrared light (with a typical wavelength between 650nm and 950nm) through the scalp and skull [3-5], measures the concentration changes of oxy hemoglobin (HbO), deoxy hemoglobin (Hb) and total hemoglobin (HbT) in the superficial cortical regions of the brain. fNIRS has high temporal resolution and reasonable spatial resolution, and thus provides an effective way to measure the resting state functional connectivity (RSFC) of adults and neonate [6-8]. Functional connectivity, presented as slow spontaneous oscillations (<0.1Hz, also known as low frequency fluctuation, LFF) during the resting state, was found by functional magnetic resonance imaging (fMRI) in typical brains [9, 10].

Autism Spectrum Disorder (ASD) is characterized by impaired social interactions, communication deficits and restricted, repetitive pattern of interests and behaviors [11]. Autism has been defined as a neuro-developmental disorder of synchronization of neural activity. Moreover, disrupted neural synchronization emerges very early (12-24 months), which could be an important early diagnostic mark for autism [12]. fMRI studies have revealed that autism spectrum disorder is linked to an abnormal pattern of brain functional connectivity, with or without tasks [13-15]. However, most of these pieces of evidence were provided from studies of adults or adolescents, mostly because fMRI has limits for child and infant studies, especially for awoken autistic children.

fNIRS have been previously used for studying autistic children under certain cognitive tasks [16, 17]. So far there is no report of fNIRS study on the RSFC in autistic children. In this paper we propose to use fNIRS as a cheaper and easy-to-operate neuro-imaging technique to find some characteristic features of RSFC neural activity of autistic children.

**2. Method**

*2.1 Participants and protocol*

In order to compare the patterns of RSFC between autistic and normal children, we recruited 10 children (*mean age* = 9.0 ± 1.3) who were diagnosed as ASD and 10 controls (*mean age* =8.9 ± 1.4) for our study. All of the participants were boys and right-handed. During the experiment, children sat in a silent room with dim light. They were asked to close their eyes and minimize movement for 8 minutes. After a participant kept stable for 2 minutes, we began to record the signals. The experiment protocol was approved by the academic ethics committee of South China Normal University and the Institutional Review Board of Guangzhou Rehabilitation and Research Center for Children with ASD, where the experiment was carried out. Parents of each child have signed the informed consent.

*2.2 Experiment setup*

44 channels of an fNIRS system (FOIRE-3000, Shimadzu Corporation, Kyoto, Japan) were used to assess the LFFs of neural activity of the left and right inferior frontal cortices (IFC) and temporal cortices (TC) of the brain. The absorptions of three wavelengths (780nm, 805nm, and 830nm) of near infrared light were measured with a sampling rate of 14.286 Hz and then transformed into concentration changes of HbO, Hb and HbT by the modified Beer-Lambert law [8]. The international 10-10 system [18] was adopted to locate the TC and IFC. As shown in Fig. 1 (a), (b) and (c), there were 22 channels in each hemisphere covering IFC and TC. The source-detector distance was fixed at 3 cm. Before the experiment starts, we ensured that all 44 channels worked properly after the optical probes were fixed on the helmet. If there was a channel warning in the adjustment procedure of the system, we tried to repair it by removing hair occultation or making better contact between probes and the scalp.

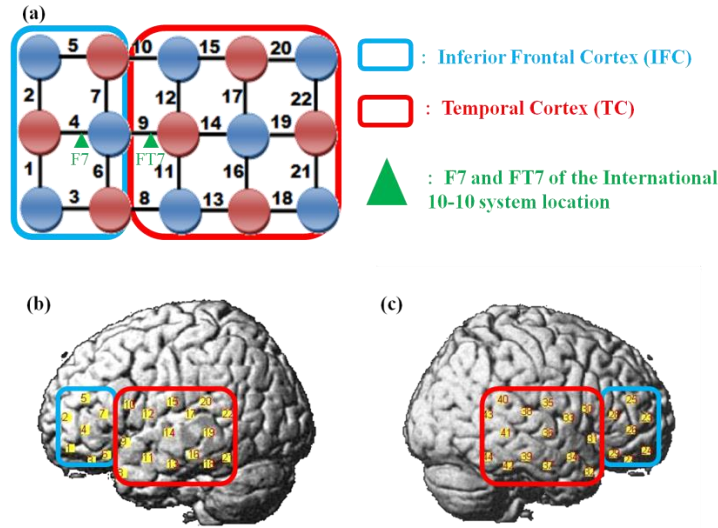

Fig.1. (a) The position of the optical probes (7 sources: red circles, 8 detectors: blue circles) in the left hemisphere. A black line connecting a source and a detector presents a data channel, which has a number alongside. 22 channels were used to cover both inferior frontal cortex (IFC) and temporal cortex (TC, including superior temporal gyrus and middle temporal gyrus). There are 7 channels (left hemisphere: from Channel 1 to Channel 7, right hemisphere: from Channel 23 to Channel 29) covering IFC and 15 channels (left hemisphere: from Channel 8 to Channel 22, right hemisphere: from Channel 30 to Channel 44) covering TC. IFC was boxed with the blue line and TC was boxed with the red line in (a), (b) and (c). Channels 4 and 9 were located in the F7 and FT7 in the international 10-10 system, respectively. The settings of the optical probes in the right hemisphere were identical with those of the probes in the left hemisphere through the anatomical symmetry. (b) Channels' numbers were marked on the left hemisphere of a phantom brain. (c) Channels' numbers were marked on the right hemisphere of a phantom brain.

*2.3 Data analysis*

Before calculating the time course correlation values and generating correlation maps, data of each channel of each individual was preprocessed in two steps. First, in order to remove possible physiological noise and signals beyond our study, a band pass filtered between 0.009-0.08 Hz [6-9] was used. Secondly, the global signal that was caused by respiration or blood pressure [8], estimated by averaging the time series over all channels, was regressed out for each channel. This is a standard procedure in fMRI's functional connectivity analysis [12]. The global signal may cause some overestimation of the connectivity if not regressed out [8].

For the correlation analysis, we calculated the Pearson correlation coefficient *r* between the time course of each channel and the corresponding channel in the other hemisphere from the region of interest (ROI; e.g., overall, IFC and TC in this study). For correlation maps, we chose a channel from a ROI as a seed and calculated *r* between the time course of the seed and the time course of all other channels in the bilateral ROI. The position of the seed was defined as the first channel in the ROI in the left hemisphere (Channel 1 in IFC and Channel 8 in TC). The correlation value between the seed channel and other channels was mapped onto the channel geometry. Then, we use interpolation to fill the blank area between the adjacent channels. The average correlation value of the autism/control group was estimated by converting *r* values to *z* values with Fisher's r-z transformation for each participant, and then converting back to the averaged *z* value to obtain the averaged *r* value.

## 3. Results

*3.1 Interhemispheric correlation*

Results of the correlation analysis are shown in Fig. 2. We use a Student's t-test [19] to study the statistical significance (*p* value) of the difference between the average correlation values for autistic and normal children. If the *p* value obtained from the t-test is lower than 0.05, we can infer that the difference is significant (smaller *p* value corresponds to a "more meaningful" result). Our result of the t-test indicated that children with autism exhibited significantly smaller overall left-right correlation values of HbO than the controls ($r_{autism} = 0.144 \pm 0.085$, $r_{normal} = 0.318 \pm 0.194$, $p < 0.05$). Those with autism had weaker overall interhemispheric functional connectivity than the controls (including all the channels we used, covering both IFC and TC). In particular, autism showed significantly weaker interhemispheric functional connectivity ($r_{autism} = 0.112 \pm 0.098$, $r_{normal} = 0.318 \pm 0.163$, $p < 0.01$) in TC (HbO), a cortical area associated with auditory sensation and language processing. The HbO value in IFC and the other two indicators of hemodynamic response (Hb and HbT) in all ROIs in this study did not show significant difference between the autistic children and the controls ($p > 0.05$). The exact *p* value for each statistical analysis was show in red in Fig. 2.

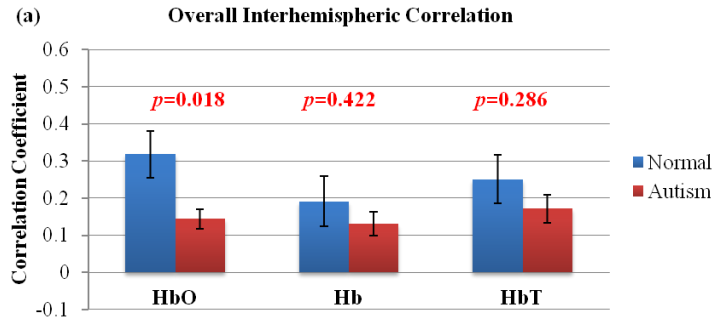

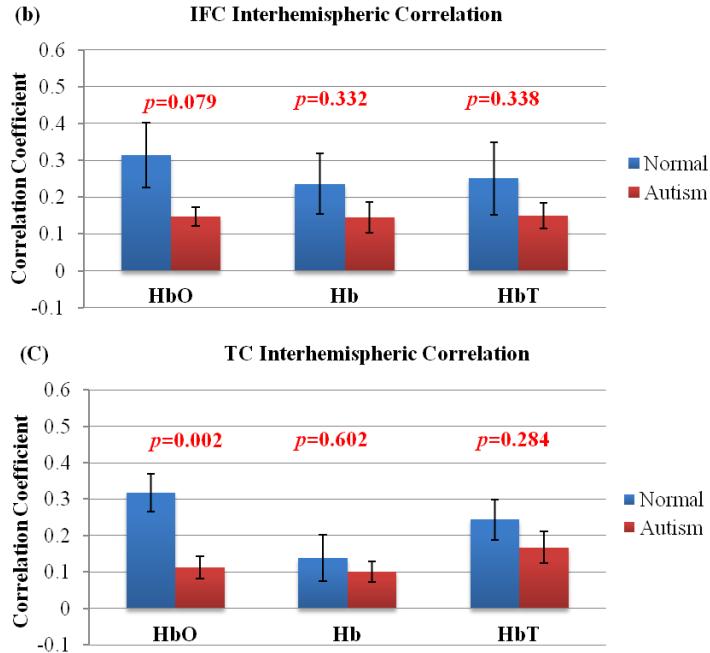

Fig. 2. Interhemipheric correlation in ROIs of children with autism (red, $n = 10$) and controls (blue, $n = 10$). Error bars are standard error of mean across participants. (a) Overall interhemipsheirc correlations (including all channels in each hemisphere) of HbO, Hb and HbT for autistic children and controls. The autism group shows significantly weaker overall left-right correlation ($p = 0.018$) than controls in HbO. (b) Interhemipsheirc correlations of HbO, Hb and HbT in IFC for autism and control group. (c) Interhemipsheirc correlations of HbO, Hb and HbT in TC for autism and control group. Autistic children exhibited weakest left-right correlation in the TC area in terms of indicator HbO.

*3.2 Correlation maps*

For comparison, Fig. 3 gives correlation maps in the TC area for a normal child and an autistic child. All normal children displayed a similar correlation pattern with strong left-and-right symmetry; however, the correlation maps for those with autism are not so similar in pattern, but share very poor left-and-right symmetry.

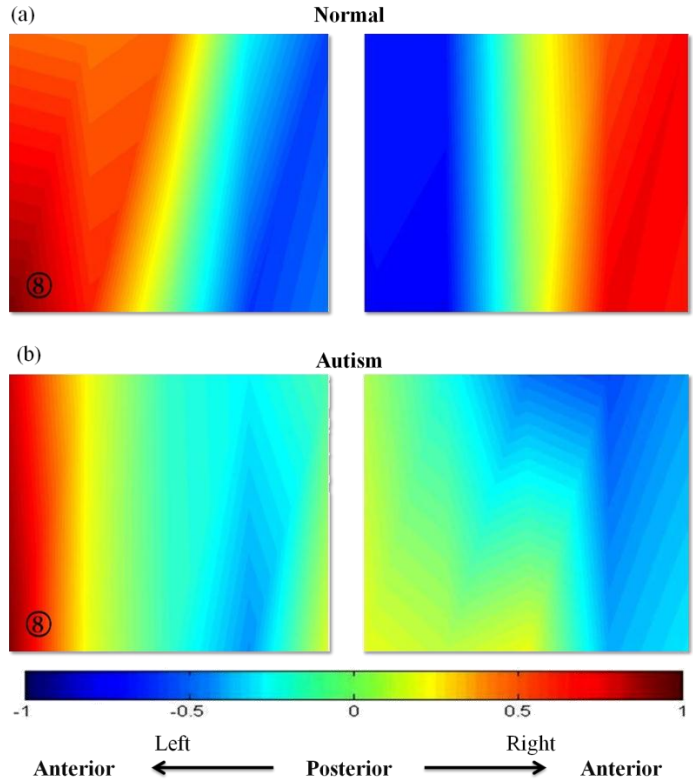

Fig.3. HbO correlation maps in the temporal cortex (TC) for a normal child and a typical autistic child. The selected seed (Channel 8, see Fig. 1(a) and (b)) was marked on the map. (a) HbO correlation maps for a normal child in TC (interhemispheric correlation value is 0.530). (b) HbO correlation maps for a autistic child in the TC (interhemispheric correlation value is 0.036). Colors represent the strength of the correlation with the seed.

## 4. Conclusions

In summary, by using fNIRS for studying RSFC of autism children, for the first time, we observed reduced interhemispheric functional connectivity in autism children. Compared with the controls, children with autism showed significantly lower interhemispheric correlation (HbO) in overall ROIs, especially in the TC. In correlation maps (HbO), autism did not have a symmetric pattern like the normal group. For the three hemodynamic parameters measured by fNIRS, HbO provided the best sensitivity to reflect the divergence of connectivity of different groups (autism vs. normal), suggesting that this contrast may be used to identify the disrupted neural connectivity of autism. Moreover, our findings of significantly lower functional connectivity in the temporal cortex are in line with fMRI studies with adolescents, adults and toddlers with autism [12, 14, 15], suggesting that fNIRS is a much cheaper and yet effective technique to detect the abnormality of neural activity of awoken children with autism.

**Acknowledgments**

This work was supported by Guangdong Innovative Research Team Program (No. 201001D0104799318), the National Basic Research Program (973) of China (2011CB503700), the Swedish Research Council and SOARD. We thank Prof. Jun Li in Centre for Optical and Electromagnetic Research of SCNU for helping with data analysis and imaging processing. We also thank Profs. Heyong Shen and Lan Gao of SCNU for discussion, and Zhifang Zhu, Lina Qiu, Xinge Li and Wei Cao for experimental helps.